\documentclass[prd,preprintnumbers,floatfix,aps,nofootinbib,notitlepage,showpacs,twocolumn,superscriptaddress]{revtex4}
\usepackage{amsmath}
\usepackage{graphicx}
\usepackage{epstopdf}
\usepackage{latexsym}
\usepackage{amssymb}

\newcommand\kasq{\ensuremath{\kappa^2}}
\newcommand\ka{\ensuremath{\ka}}
\newcommand\veff{\ensuremath{V_{\textrm{eff}}}}

\newcommand\req[1]{Eq.~(\ref{#1})}

\begin{document}

\title{Screening three-form fields}

\author{Tiago Barreiro}
\email{tmbarreiro@ulusofona.pt}
\affiliation{Departamento de Matem\'atica,  ECEO, Universidade Lus\'ofona de Humanidades e 
Tecnologias, Campo Grande, 376,  1749-024 Lisboa, Portugal}
\affiliation{Instituto de Astrof\'isica e Ci\^encias do Espa\c{c}o, Universidade de Lisboa,
Faculdade de Ci\^encias, Campo Grande, PT1749-016 Lisboa, Portugal}

\author{Ugo Bertello}
\affiliation{Instituto de Astrof\'isica e Ci\^encias do Espa\c{c}o, Universidade de Lisboa,
Faculdade de Ci\^encias, Campo Grande, PT1749-016 Lisboa, Portugal}

\author{Nelson J. Nunes}
\email{njnunes@fc.ul.pt}
\affiliation{Instituto de Astrof\'isica e Ci\^encias do Espa\c{c}o, Universidade de Lisboa,
Faculdade de Ci\^encias, Campo Grande, PT1749-016 Lisboa, Portugal}

\begin{abstract}
Screening mechanisms for a three-form field around a dense source such as the Sun are investigated. Working with the dual vector, we can obtain a thin-shell where field interactions are short range. The field outside the source adopts the configuration of a dipole which is a manifestly distinct behaviour from the one obtained with a scalar field or even a previously proposed vector field model. 
We identify the region of parameter space where this model satisfies present solar system tests.
\end{abstract}
\pacs{xxx}

\maketitle

\section{Introduction}
In the last one hundred years, the  theory of General Relativity has given us a framework 
to explore Gravity from the orbits of planets to the evolution of the whole Universe. 
Nonetheless, a number of puzzling issues still remain, the most disconcerting one being the lack of explanation for the recent acceleration of the Universe.
 When seeking to extend General Relativity, theorists often invoke the presence of a dynamical scalar field either in the gravitational 
part of the action or instead, in the matter sector. This procedure is naturally dangerous as the field might couple to the rest of the world leading to fifth-forces and violating the equivalence principle.
In order to avoid such constraints, screening mechanisms that impose a 
constant value of the field inside a body 
such as the Earth or the Sun, have been proposed. 
The way this is achieved consists in choosing  a particular form for the potential and coupling of the field with baryons that ensures a large mass in the interior of the body, hence imposing short range interactions. In space, however, the field's mass is small and mediates an interaction of gravitational strength.  These models are known by the name of chameleon mechanism \cite{Khoury:2003aq,Khoury:2003rn,Gubser:2004uf,Brax:2004qh,Mota:2006ed,Mota:2006fz,Brax:2010kv,Khoury:2013yya}.
An alternative set up is the symmetron \cite{Hinterbichler:2010es}, with a large vacuum expectation value in environments of low density and a small expectation value in environments of high density. As the coupling is proportional to the vacuum expectation value, the field decouples from the matter fields in regions of high density. 
Derivative couplings \cite{Noller:2012sv,Koivisto:2012za} have also been proposed as screening mechanisms. In the Vainshtein mechanism 
\cite{Vainshtein:1972sx,ArkaniHamed:2002sp,Deffayet:2001uk,Babichev:2013usa}, the derivative self-couplings of the scalar field become large in the neighbourhood of a massive body. 
These non-linear contributions increase the kinetic terms of perturbations and suppress the strength of the interactions of the scalar field with matter.  

 Vector fields also exist in nature and one naturally wonders whether screening mechanisms also exist for those. This was investigated in Ref.~\cite{BeltranJimenez:2013fca} where it is shown that the mechanism is very similar to the symmetron and indeed such features exist with the additional effect that Lorentz invariance can also be shielded for a vector field. We can take this investigation one step further and study how a higher rank tensor behaves inside and in the neighbourhood of a compact object. In this article, we will focus on a vector dual to a three-form field \cite{Koivisto:2009ew,Koivisto:2009fb,Koivisto:2012xm}. The equations of motion are distinct from the ones of the vector field in Ref.~\cite{BeltranJimenez:2013fca} and consequently, the phenomenology obtained also differs considerably. Furthermore, we choose a form of the potential and coupling that results in a large vacuum expectation value of the field in high dense regions and conversely a small value in low dense regions. This allows us to recover homogeneity and isotropy  on large scales.  We first motivate the vector model from a three-form action and obtain the equations of motion.  With our specific choice of potential and coupling, the field profile around a spherical source is computed and compared against previous scalar and vector field model solutions. Finally we place bounds on the model parameters from current observational limits.

%%%%%%%%%%%%%%%%%%%%%%%%%%%%%%%%%%%%%%%%%%%%%%%%%%
\section{Coupled three-form theory}

We start from the action for a three-form field $A$ minimally coupled to gravity with a conformally coupled matter Lagrangian density $\tilde{\mathcal{L}}_M$,
\begin{align}
S = \int d^4 x \sqrt{-g} \left[ \frac{1}{2 \kasq} R - \frac{1}{48} F^2 - V(A^2) \right]
+ \int d^4 x \tilde{\mathcal{L}}_m,
\end{align}
where $g$ is the determinant of the metric $g^{\mu \nu}$, $R$ is the Ricci scalar for the same metric, 
$F^2 = F^{\alpha \beta \gamma \delta } F_{\alpha \beta \gamma \delta}$ where the four-form $F$ is a generalized Faraday form given by
$F_{\alpha \beta \gamma \delta} = 4 \nabla_{[\alpha} A_{\beta \gamma \delta]}$ and $V$ is the potential for the three-form field $A$ with $A^2 = A^{\alpha \beta \gamma} A_{\alpha \beta \gamma}$. Throughout we use signature $(-,+,+,+)$.

We consider the matter sector to consist of a pressureless perfect fluid with energy density $\tilde{\rho}$. Its Lagrangian density $\tilde{\mathcal{L}}_m = \sqrt{-\tilde{g}} L_m$ depends on the metric $\tilde{g}^{\mu \nu}$ which is related to the gravity and three-form metric through a conformal transformation
\begin{align}\label{confrel}
\tilde{g}_{\mu \nu} = \Omega^2(A^2) g_{\mu \nu},
\end{align}
where we assume the conformal factor $\Omega$ to be  only a function of $A^2$.

Varying the action with respect to the three-form field yields the equation of motion
\begin{align}\label{3form-eqn-motion}
\nabla_\alpha F^{\alpha \beta \gamma \delta} =
12 \left(\frac{\partial V}{\partial A^2} + \rho \frac{\partial \Omega}{\partial A^2}\right)
A^{\beta \gamma \delta},
\end{align}
where  we have defined the usual matter energy density in the Einstein frame as $\rho = \Omega^3 \tilde{\rho}$.

We can look at this as having an effective potential $V(A^2) + \rho \Omega(A^2)$ with an explicit dependence on the surrounding fluid energy density.

Since $F^{\alpha \beta \gamma \delta}$ is antisymmetric, we also have the constraint equation
\begin{align} \label{3form-constraint}
\nabla_\beta \nabla_\alpha F^{\alpha \beta \gamma \delta} = 0.
\end{align}

%%%%%%%%%%%%%%%%%%%%%%%%%%%%%%%%%%
\section{Dual vector field}
For ease of calculation it is convenient to recast the theory using a vector field.  For this purpose we introduce the Hodge dual forms of $A$ and $F$, such that 
\begin{align}
B_\alpha &= \frac{1}{3!} \epsilon_{\alpha \beta \gamma \delta} A^{\beta \gamma \delta}, &
\Phi &= \frac{1}{4!} \epsilon_{\alpha \beta \gamma \delta} F^{\alpha \beta \gamma \delta}, \\
\intertext{with the inverse relations}
A^{\beta \gamma \delta} &= \epsilon^{\beta \gamma \delta \alpha} B_{\alpha}, &
F^{\alpha \beta \gamma \delta} &= -\epsilon^{\alpha \beta \gamma \delta} \Phi,
\end{align}
where $\epsilon$ is the fully antisymmetric Leci-Civita tensor\footnote{$\epsilon_{\alpha \beta \gamma \delta} = \sqrt{-g}\, \varepsilon_{\alpha \beta \gamma \delta}$,
with $\varepsilon$ the Levi-Civita symbol.}.
It is easy to verify that $\Phi = \nabla^\alpha B_{\alpha}$, and also that
\begin{align}
A^2 &= -6 B^2, &
F^2 &= -24 \Phi^2.
\end{align}
The equation of motion (\ref{3form-eqn-motion}) can now  be written as
\begin{align} \label{1form-eqn-motion}
\nabla_\alpha \Phi =
-2 \left(\frac{\partial V}{\partial B^2} + \rho \frac{\partial \Omega}{\partial B^2}\right)
B_\alpha,
\end{align}
and the constraint equation (\ref{3form-constraint}) becomes
\begin{align}\label{Bconstraint}
\nabla_\alpha \nabla_\beta \Phi - \nabla_\beta \nabla_\alpha \Phi = 0.
\end{align}
Defining an effective potential as $\veff(B^2) = - V(B^2) - \rho \Omega(B^2)$, the equation of motion (\ref{1form-eqn-motion}) simply yields
\begin{align}\label{Bevol}
\nabla_\alpha \Phi = \frac{\partial \veff}{\partial B^{\alpha}}.
\end{align}
Recalling that $\Phi = \nabla^\mu B_\mu$, we can write the equation of motion \req{Bevol} in terms of the vector field  only 
\begin{align}\label{Bevol2}
\nabla_\alpha \left( \nabla^\mu B_\mu \right) = \frac{\partial \veff}{\partial B^{\alpha}}.
\end{align}
Note that this is fairly different from Ref.~\cite{BeltranJimenez:2013fca} where the vector field equation is given by
$\Box B_\mu = \frac{\partial \veff}{\partial B^{\mu}}$. 

%%%%%%%%%%%%%%%%%%%%%%%%%%%%%%%%%
\section{Screening mechanism}
%%%%%%%%%%%%%%%%%%%%%%%%%%%%%%%
\subsection{Minima}

We want to look at the field around massive astrophysical objects. We can model this using a presureless spherically symmetric object with a homogeneous energy density $\rho_c$ and radius $r_c$ surrounded by a low and homogeneous energy density background $\rho_b$.

We are interested in static solutions in Minkowski space, therefore  all our time derivatives are set to zero. From \req{1form-eqn-motion} we can see that, provided $\partial V_{\rm eff} /\partial B^2 \neq 0$, the time component of $B_\mu$ has to be zero, that is $B_\mu = (0,\vec{B})$. Also, the dual scalar field $\Phi$ is the divergence of the spatial vector, $\Phi =  \vec{\nabla} \cdot \vec{B}$. Hence, we will only work in 3 dimensions in what follows.

The first derivative of the effective potential obtained from equation \req{1form-eqn-motion} is
\begin{align}\label{effpot-der1}
\frac{\partial \veff}{\partial B^i} =  -2 \left( \frac{\partial V}{\partial B^2} + \rho \frac{\partial \Omega}{\partial B^2} \right)  B_i \;.
\end{align}
and the equation of motion for $\vec{B}$ is
\begin{align}\label{3dBevol}
\vec{\nabla} \left( \vec{\nabla} \cdot \vec{B} \right) = \frac{\partial \veff}{\partial \vec{B}} \; .
\end{align}

From \req{effpot-der1} we find that a minimum of the effective potential can occur for $\vec{B} = 0$ but also whenever the combination $\frac{\partial V}{\partial B^2} + \rho \frac{\partial \Omega}{\partial B^2}$ vanishes, yielding a broken $O(3)$ symmetry minimum. In either instance, the effective masses come from the second derivative of the effective potential,
\begin{multline}
\frac{\partial \veff}{\partial B^i \partial B^j}
= -2 \left( \frac{\partial V}{\partial B^2} + \rho \frac{\partial \Omega}{\partial B^2} \right) \delta_{ij}
\\
- 4 \left( \frac{\partial^2 V}{(\partial B^2)^2} + \rho \frac{\partial^2 \Omega}{(\partial B^2)^2} \right) 
B_i B_j \; .
\end{multline}

At $\vec{B} = 0$ this yields an effective mass matrix
\begin{align}
m^2_0~ \delta_{ij} = \left.  \frac{\partial \veff}{\partial B^i \partial B^j} \right|_{\vec{B}=0}
= -2  \left( \frac{\partial V}{\partial B^2} + \rho \frac{\partial \Omega}{\partial B^2} \right)\delta_{ij} \;.
\end{align}
When the symmetry is  broken, we get a displaced minimum at the critical value $\vec{B} = \vec{B}_c$ such that $\left. \frac{\partial V}{\partial B^2} + \rho \frac{\partial \Omega}{\partial B^2} \right|_{\vec{B}_c} = 0$. In this minimum the field will get a mass in the $\vec{B}_c$ direction
\begin{align}
m_c^2 =  \left. -4 \left( \frac{\partial^2 V}{(\partial B^2)^2} + \rho \frac{\partial^2 \Omega}{(\partial B^2)^2} \right) \right|_{\vec{B}_c} B_c^2 \; ,
\end{align}
whereas in the other two spatial orthogonal directions it will have an effective mass equal to zero.

\begin{figure}
\includegraphics[width=8cm,angle=0]{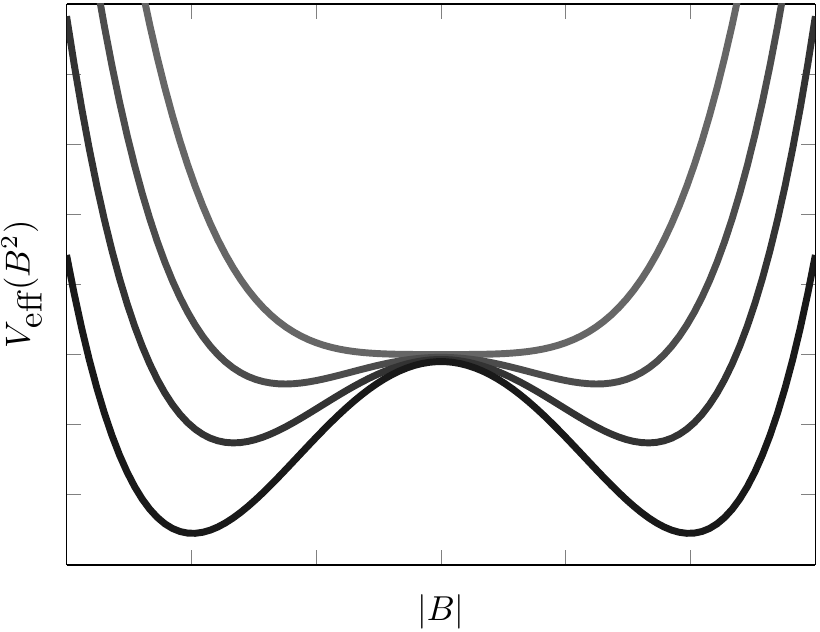}
\caption{The effective scalar potential. Darker lines correspond to a larger local energy density. The scalar potential develops non-zero minima for larger energy densities, with an effective mass $m_s^2$ larger than the effective mass at zero $m_0^2$.}
\label{potplot}
\end{figure}

From now on we will use a specific form for our scalar potential
\begin{align}
V(B^2) = -\frac{1}{2} m^2 B^2 - \frac{1}{4} B^4,
\end{align}
and for our conformal coupling
\begin{align}
\Omega(B^2) = 1 + \frac{1}{2\mu^2}B^2 \; ,
\end{align}
where $m$ and $\mu$ are our only  two mass parameters which we choose to be real and positive.

The broken symmetry minimum is given by
\begin{align}\label{Bcval}
B_c^2 =  \frac{\rho}{\mu^2} \left(1 - \frac{\mu^2 m^2}{\rho} \right) ,
\end{align}
and for consistency, symmetry can be broken only when $B_c^2>0$, hence, for $\rho > \mu^2 m^2$.
Consequently, in this model we have a broken symmetry inside the massive body where the energy density $\rho_c$ is large, whereas symmetry is restored outside where the background energy density $\rho_b$ is small (see Fig.~\ref{potplot}). With this property we ensure that homogeneity and isotropy is recovered on the large scales.
Note that this is the opposite choice to the one taken in Ref.~\cite{BeltranJimenez:2013fca} where the vector symmetry is restored within the body and broken in the background.

Inside the body we assume $\rho_c \gg \mu^2 m^2$, the minimum is at $B_c^2 \approx \rho_c/\mu^2$ and the non-zero effective mass in the $\vec{B}_c$ direction is
\begin{align}\label{mc2}
m_c^2 = 2  \frac{\rho_c}{\mu^2} \left(1 - \frac{\mu^2 m^2}{\rho_c} \right)
\approx 2 \frac{\rho_c}{\mu^2} \; .
\end{align}

Outside the body we  have a small energy density $\rho_b \ll \mu^2 m^2$ and the minimum is at $B = 0$ with an effective mass
\begin{align}\label{m02}
m_0^2 = m^2 \left( 1 - \frac{\rho_b}{\mu^2 m^2} \right)
\approx m^2 \; .
\end{align}
Again, assuming that we have a massive  body where $\rho_c \gg \mu^2 m^2$, we can see from Eqs.~(\ref{mc2}) and (\ref{m02}), that $m_c^2 \gg m_0^2$.

%%%%%%%%%%%%%%%%%%%%%%%%%%%%%%%%%%%%%%
\subsection{Field profile}
%%%%%%%%%%%%%%%%%%%%%%%%%%%%%%%%%%%%%%
\subsubsection{Core region}

We can now look for solutions around the minima of the potential.
Inside the massive body with energy density $\rho_c$
symmetry is broken and the minimum of the potential is at $\vec{B}_c$. The field only acquires a mass in the $\vec{B}_c$ direction which we can identify with the $z$ direction. 
In \req{3dBevol} only the $B_z$ component gets a non-trivial solution and the linear perturbations in this direction is proportional to $e^{\pm m_c z}$. As a first approximation we assume $m_c z \ll 1$ and therefore neglect this correction such that at the body core the field is constant, $\vec{B} = \vec{B}_c$. It is convenient to rewrite this solution as the gradient of a scalar field $\vec{B} = \vec{\nabla} \phi$, with 
\begin{align}\label{ficore}
\phi_\textrm{c} = B_c \, r \cos \theta,
\end{align}
where $z = r \cos \theta$, $r$ being the radius and $\theta$ the polar angle in spherical coordinates.

%%%%%%%%%%%%%%%%%%%%%%%%%%%%%%%
\subsubsection{Shell region}

As the field approaches the boundary, it reaches an intermediary region where the field becomes dislocated from its minimum and the approximation used around the core minimum no longer holds. This takes place at an inner shell radius $r_s$. 
In this shell region between $r_s$  and the body radius $r_c$, the effective potential is dominated by the contribution from the matter coupling $-\rho \Omega(B^2)$. Thus, neglecting the scalar potential,  \req{3dBevol}  becomes
\begin{align}\label{3deq-shell}
\vec{\nabla} \left (\vec{\nabla} \cdot \vec{B} \right) = - m_s^2 \vec{B},
\end{align}
where $m_s^2 = 2 \rho_c \frac{\partial \Omega}{\partial B^2} = \rho_c/\mu^2$.

Since $m_s^2$ is a constant, from \req{3deq-shell} we see that  $\vec{B}$ can be written as the gradient of a scalar field, $\vec{B} = \vec{\nabla} \phi$ (this could also be seen from the constraint equation (\ref{Bconstraint}), in fact, $\Phi = -m_s^2 \phi$). We can now rewrite \req{3deq-shell} in terms of $\phi$. Integrating and discarding the arbitrary integration constants, we get the field equation for $\phi$ 
\begin{align}\label{eq-phi-shell}
 \nabla^2 \phi + m_s^2 \phi = 0.
\end{align}

 In spherical coordinates with azymuthal symmetry around the $z$ direction, we can separate the variables and expand the solution as
\begin{align}
\phi = \sum_{l} f_l(r) P_l(\cos \theta),
\end{align}
where $P_l(\cos \theta)$ are  Legendre polynomials. Since we have $m_s^2>0$, the $f_l(r)$ functions are  the spherical Bessel functions, $j_l(m_s r)$ and $y_l (m_s r)$. 
Recall that $P_1(\cos \theta) = \cos \theta$, consequently, comparing this solution with
the core solution \req{ficore}   and imposing boundary conditions, only the $l=1$ 
solution takes non-zero coefficients\footnote{Note that the $l=0$ solution corresponds to the usual symmetron solution for a scalar field \cite{Hinterbichler:2010es} or to the vector one in \cite{BeltranJimenez:2013fca}.}. The solution in the shell region is therefore given by
\begin{align}
\phi_\textrm{s} = \big( a j_1(m_s r) + b y_1(m_s r) \big) \cos \theta,
\end{align}
where $a,b$ are constants and the $l=1$ spherical Bessel functions are 
\begin{align}
j_1(m_s r) &= \frac{\sin(m_s r)}{(m_s r)^2} - \frac{\cos(m_s r)}{m_s r},
\\
y_1(m_s r) &=  - \frac{\cos(m_s r)}{(m_s r)^2} - \frac{\sin(m_s r)}{m_s r}.
\end{align}

%%%%%%%%%%%%%%%%%%%%%%%%%%%%%%
\subsubsection{Outside region}

Finally, outside the body we have a small energy density $\rho_b$ and the minimum of the potential is at 
$\vec{B} = 0$ with an effective mass $m_0$. To linear order \req{3dBevol} becomes
\begin{align}\label{3deq-out}
\vec{\nabla} \left (\vec{\nabla} \cdot \vec{B} \right) =  m_0^2 \vec{B}.
\end{align}
Again we can write our field as a  gradient $\vec{B} = \vec{\nabla} \phi$,  where $\phi$ obeys the equation
\begin{align}\label{eq-phi-out}
 \nabla^2 \phi - m_0^2 \phi = 0.
\end{align}
The solution is  again  an expansion $\phi = \sum_l f_l(r) P_l(\cos \theta)$, however, now with a negative mass term meaning that the $f_l(r)$ functions turn out to be the modified spherical Bessel functions which for $l=1$ are
\begin{align}
i_1(m_0 r) & = \frac{(m_0 r \cosh(m_0 r) - \sinh(m_0 r))}{(m_0 r)^2}, \\
k_1(m_0 r) &  = \frac{(m_0 r+1)e^{-m_0 r}}{(m_0 r)^2} .
\end{align}
Since we want our field outside the body to be finite at large $r$, the $i_1(m_0 r)$ coefficient has to be zero and  the outside solution is
\begin{align}
\phi_{\textrm{out}} = c \, k_1(m_0 r) \cos \theta,
\end{align}
where $c$ is a constant.
\subsubsection{Global profile}

Altogether our global field profile is $\vec{B} = \vec{\nabla} \phi$ with the $\phi$ field given by the $l=1$ solutions, $\phi = f(r) \cos \theta$, and the piecewise function $f(r)$  given by
\begin{align}\label{fofr}
f(r) = 
\begin{cases}
B_c r , & \textrm{ if } r<r_s \\
a j_1(m_s r) + b y_1(m_s r), & \textrm{  if } r_s < r < r_c \\
c \, k_1 (m_0 r), & \textrm{ if } r > r_c 
\end{cases}
\end{align}
The constants $a$, $b$ and $c$ are to be determined by imposing suitable boundary conditions.

In the case of a massive body we have $m_s r_c > m_s r_s \gg 1$ and the profile $f(r)$ in the shell, $r_s<r<r_c$, is approximately
\begin{align*}
a j_1(m_s r) + b y_1(m_s r)  \simeq - a \frac{\cos(m_s r)}{m_s r} - b \frac{\sin(m_s r)}{m_s r} \;.
\end{align*}

\begin{figure}
\includegraphics[width=8cm]{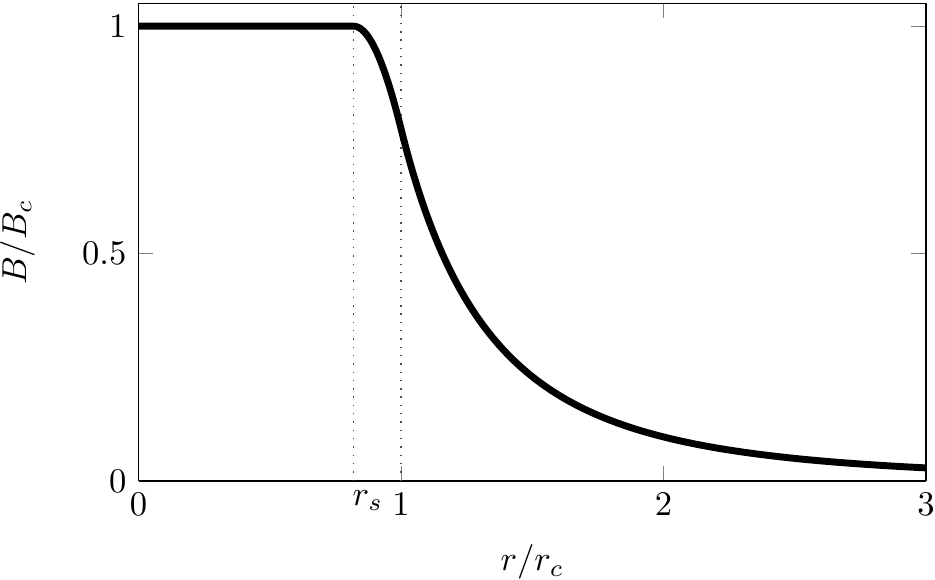} 
\caption{\label{fieldprof} $B(r)$ profile  in the $\theta = 0$ direction. In this direction the $B_\theta$ component vanishes and the field profile is continuous.}
\end{figure}

\begin{figure}
\includegraphics[width=8cm]{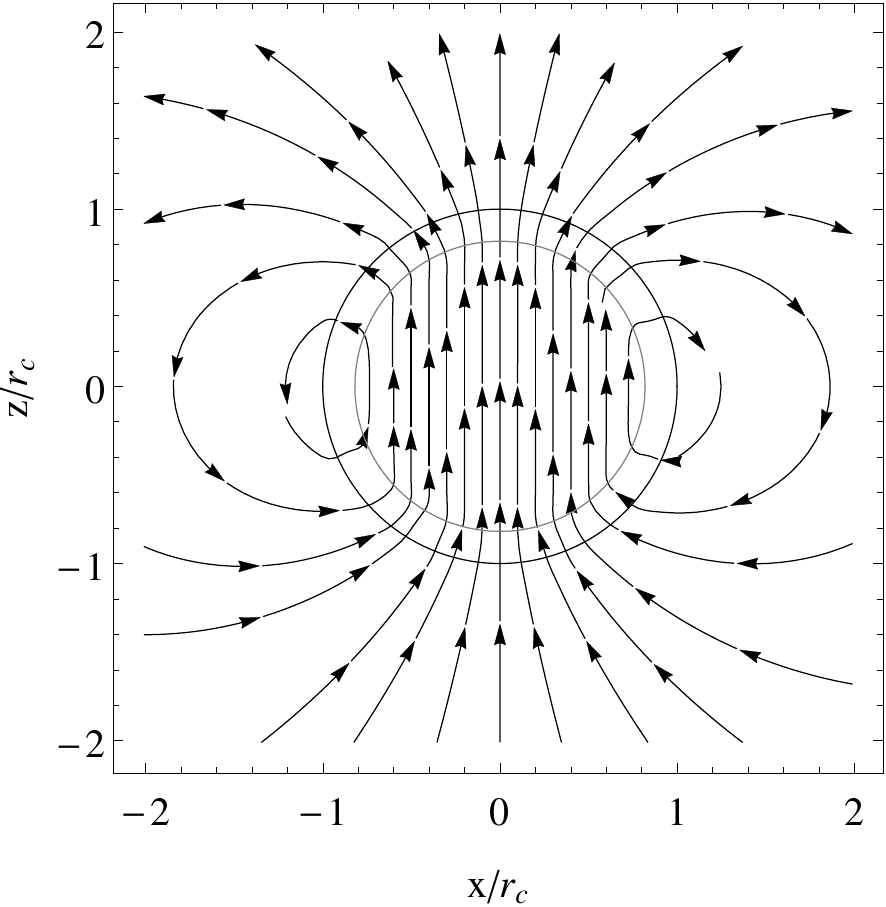}
\caption{\label{vecstream} Stream lines for the vector field $\vec{B}$. In the core, the solution is the constant vector $B_c$ aligned with the $z$ axis. The grey inner circle shows the thin shell boundary $r_s$. 
Note that the analytical approximation obtained in the text is not valid for this choice of shell thickness.}
\end{figure}

At the boundaries we expect the radial component of $\vec{B}$ to be continuous, however because of the discontinuity in the effective masses we recquire a discontinuity in the angular component of $\vec{B}$. Therefore we impose as boundary conditions continuity in the radial component and its first $r$ derivative. In practice, this corresponds to imposing that $f'(r)$ and $f''(r)$ are continuous. With these boundary conditions imposed at $r=r_s$ and $r=r_c$ we obtain
\begin{align}
a &= r_c B_c \cos(m_s r_c), &
c &= -\frac{1}{2} m_0^2 r_c^3  B_c,  \label{ac}\\
b &= -r_c B_c \sin(m_s r_c), &
r_s &= r_c - \frac{3}{m_s^2 r_c},  \label{brs}
\end{align}
where we have  assumed $m_0 r_c \ll 1$, yielding $k_1(m_0 r) \simeq 1/(m_0 r)^2$.
In Fig.~\ref{fieldprof} we show the field profile along the $z$ direction. In Fig.~\ref{vecstream} we illustrate the stream lines of the  vector field $B$ in the $(z,x)$ plane.

In the limit when $\rho_c \gg \mu^2 m^2$ we recover from \req{brs} the usual thin shell scenario, where the field is only displaced from its minimum in a thin shell inside the body.
For this model the relative shell thickness is given by $\Delta r/r_c = 3/m_s^2 r_c^2$,
the effect being stronger for larger values of $m_s r_c$.

%%%%%%%%%%%%%%%%%%%%%%%%%%%%
\section{Observational constraints}

In this section we impose bounds on the model parameters using a set of solar system data.
Since we want symmetry to be restored in our galaxy, we need $m^2\mu^2 > \rho_\textrm{gal} \simeq 10^{-30}$MeV$^4$. On the other hand, for symmetry to be broken inside the sun we require  that $m^2 \mu^2 < \rho_\textrm{sun} \simeq 6 \times 10^{-6}$MeV$^4$.

Measurements of the frequency shift in a radio signal passing close to the sun give us a constraint on the post-Newtonian parameter $\gamma$. In the Jordan frame we have the metric $\tilde{g}_{ij} = (1-2 \gamma \Phi_J) \delta_{ij}$ with $\tilde{g}_{00} = -(1+2 \Phi_J)$, where $\Phi_J$ is the gravitational potential in the Jordan frame. In the Einstein frame this becomes $g_{ij} = 1 - 2 \Phi_E$ and $g_{00} = -(1+2 \Phi_E)$. Using \req{confrel} we then obtain that
\begin{align}
\label{gamma}
\gamma = \frac{2 \Phi_E - (1-1/\Omega^2)}{2 \Phi_E + (1-1/\Omega^2)}
\approx 1 - \frac{B^2}{\mu^2 \Phi_E},
\end{align} 
assuming a small field, that is, $B/\mu \ll 1$.

 Since $\vec{B} = \vec{\nabla} \phi$ we have that
\begin{align}\label{B2bound}
B^2 & = B_r^2 + B_\theta^2 = f'(r)^2 \cos^2 \theta + \frac{f(r)^2}{r^2}  \sin^2 \theta ,
\end{align}
where $f$ is given by $f_\textrm{out}(r) = c k_1(m_0 r)$. To first order in $m_0 r \ll 1$ this yields
\begin{align}
B^2 \simeq  \frac{1}{4} \left( \frac{r_\textrm{sun}}{r} \right)^6 B_c^2 \left( 3 \cos^2 \theta + 1 \right).
\end{align}
The value of $B$ reaches its maximum at the poles, $\cos\theta = 1$. 
 Given that  $B_c^2 \approx \rho_\textrm{sun}/\mu^2$ from  \req{Bcval}, its value at
$r \gtrsim r_\textrm{sun}$, is given by
\begin{align}
B^2_\textrm{max} \approx B_c^2 \approx \frac{\rho_\textrm{sun}}{\mu^2}.
\end{align}
Obviously the field is expected to acquire a slightly smaller value at the boundary than its value at the core, $B_c$. In order to see this, we need to expand \req{B2bound} to higher order terms in $m_0 r$.  At the poles where $\cos \theta = 1$ and $r=r_{\rm sun}$, we obtain
\begin{align}
B_\textrm{max} = c \left[ -\frac{2}{m_0^2 r_{\rm sun}^3} + \frac{1}{3} m_0 + \cdots \right]
\simeq B_c \left[1- \frac{m_0^3 r_{\rm sun}^3}{6}\right],
\end{align}

This confirms that, within the thin-shell, the field decreases by $\Delta B/B \sim m_0^3 r_{\textrm{sun}}^3/6$, which is a third order effect in $m_0 r$.  
We can neglect this effect for the solar system observational bounds.

Note that we could have obtained the zero order solution simply by taking the field to be $B_c =$ constant inside the body up to the boundary and dropping altogether the intermediate region. Going to the next order, however, allows us to estimate the shell thickness and evaluate the parameter space where it is indeed small, leading to a screening effect.

\begin{figure}
\includegraphics[width=8cm]{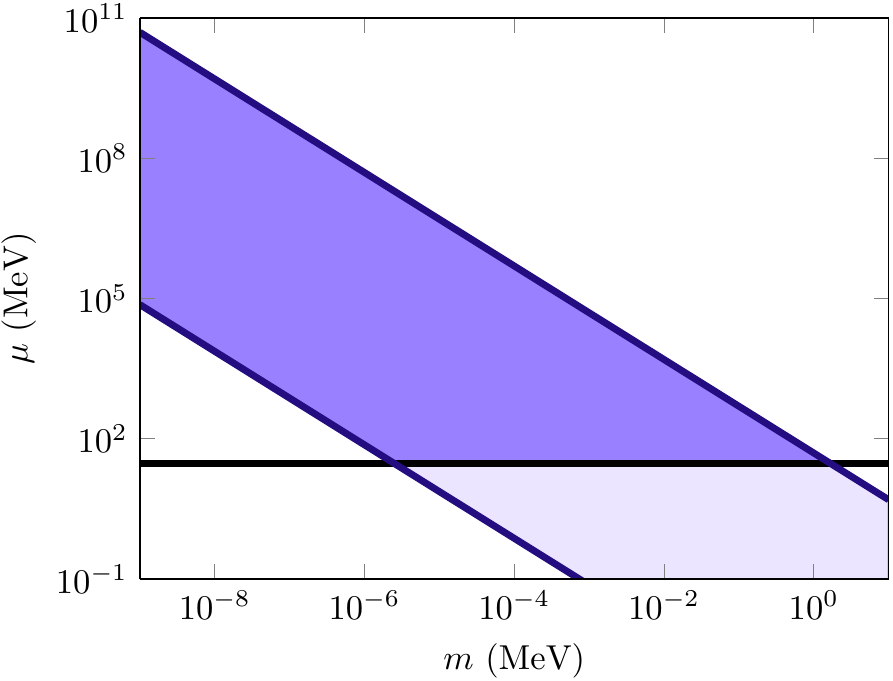}
\caption{\label{constr} Observational bounds on $m$ and $\mu$. We impose that symmetry is broken inside the sun excluding the top region and that symmetry is restored in our galaxy excluding the bottom region. The horizontal lower bound is the Cassini observational restriction obtained in \req{cassb}}
\end{figure}

The gravitational potential of the Sun at its surface is $\Phi_\textrm{sun} = G M_\textrm{sun}/r_\textrm{sun} = 10^{-6}$. Using 
\req{gamma} and  the bounds on $\gamma$ from the Cassini spacecraft measurements \cite{Bertotti:2003rm}, $|\gamma-1| \lesssim 10^{-5}$, we obtain the observational  constraint on $\mu$
\begin{align}\label{cassb}
\mu \gtrsim \left(\rho_\textrm{sun} 10^{11} \right)^{1/4} \approx 30 \, \mathrm{MeV},
\end{align} 
using $\rho_\textrm{sun} = 6 \times 10^{-6} \; \mathrm{MeV}^4$. In Fig.~\ref{constr} we show the  constraints that arise from the energy densities of the sun and of the solar system medium and the limits from Cassini on $\gamma$.

\section{Conclusions}

In this paper we have studied a three-form model with a conformal coupling to the matter sector. Working with the dual vector field, we have shown that we can implement for this model the usual screening mechanism used in theories with these types of coupling.

We have concentrated our efforts in a scenario where the vector field goes to zero at cosmological scales but where symmetry is broken inside massive bodies. This is the opposite of the choice studied in \cite{BeltranJimenez:2013fca} and not surprisingly the solutions obtained are noticeably different.

In our scenario we recover the usual thin-shell configuration, where the field inside a massive body is effectively shielded from the background, with only a thin layer where the field is displaced from its minimum solution. However, in contrast to a scalar field \cite{Khoury:2003aq,Hinterbichler:2010es}, or  the vector field solution investigated in \cite{BeltranJimenez:2013fca} in the scenario presented here the vector field solution has the configuration of a dipole. This means that around a massive body the field magnitude decays as $\sim 1/r^3$ and has an angular dependence. This is a strong signature of this model that distinguishes it from others that also have a thin-shell screening mechanism.

Using the current observational constraints on the post-Newtonian parameter $\gamma$ we obtained a lower bound on the value of the conformal coupling mass $\mu \gtrsim 30 \textrm{MeV}$. This result, together with constraints on the model viability based on the energy densities of the sun and of the solar system medium, imposes an upper bound on 
 the scalar potential mass, $m$ of a few MeV.

\begin{acknowledgments}
The authors thank Philippe Brax, Carsten van de Bruck, and Jos\'e Beltr\'an Jim\'enez for comments on the manuscript.  
The authors were supported by the Funda\c{c}\~{a}o para a Ci\^{e}ncia e Tecnologia
(FCT) through the grant UID/FIS/04434/2013. 
\end{acknowledgments}

\bibliography{bibl}

\end{document}